\documentclass[
  aps,
  physrev,
  twocolumn,
  10pt,
  amsmath,
  amssymb,
  superscriptaddress,
  floatfix
]{revtex4-2}

\usepackage{xcolor}
\usepackage{package}

\pdfoutput=1

\begin{document}
\title{Enhancing Pauli Correlation Encoding for quantum optimization via systematic expressivity analysis}

\author{Riku Usuki}%
\affiliation{%
  Graduate School of Engineering Science, The University of Osaka, 1-3 Machikaneyama, Toyonaka, Osaka 560-8531, Japan
}
\author{Don Arai}%
\affiliation{%
  Graduate School of Engineering Science, The University of Osaka, 1-3 Machikaneyama, Toyonaka, Osaka 560-8531, Japan
}
\author{Ken N. Okada}%
\affiliation{%
  Center for Quantum Information and Quantum Biology,
  The University of Osaka, 1-2 Machikaneyama, Toyonaka 560-0043, Japan
}%
\author{Keisuke Fujii}%
\affiliation{%
  Graduate School of Engineering Science, The University of Osaka, 1-3 Machikaneyama, Toyonaka, Osaka 560-8531, Japan
}

\affiliation{%
  Center for Quantum Information and Quantum Biology,
  The University of Osaka, 1-2 Machikaneyama, Toyonaka 560-0043, Japan
}%
\affiliation{%
  RIKEN Center for Quantum Computing (RQC),
  Hirosawa 2-1, Wako, Saitama 351-0198, Japan
}%
\affiliation{%
  Graduate School of Informatics, Kyoto University,
  Yoshida-Honmachi, Sakyo-ku, Kyoto 606-8501, Japan
}

\date{\today}

\begin{abstract}
Quantum approaches for combinatorial optimization problems have attracted considerable attention in recent years.
Among these approaches, Pauli Correlation Encoding (PCE) has emerged as a promising framework for quantum devices with limited qubit resources because it embeds optimization variables in expectation values of Pauli strings.
However, the mechanisms underlying its performance and the reasons for its saturation remain unclear.
In this work, we investigate these questions through a systematic analysis of expressivity and trainability.
First, we compare PCE with classical surrogate models based on tensor networks whose structures progressively approach the topology of the PCE circuit.
The results show that PCE attains comparable solution quality with substantially fewer trainable parameters, indicating strong parameter efficiency.
Second, to determine whether the performance saturation of conventional PCE is caused by insufficient expressivity or by optimization difficulty, we perform a diagnostic expressivity test in which the circuit is trained toward reference configurations for Max-Cut.
The results show that even shallow PCE circuits can represent strong solutions, indicating that the main bottleneck is not the representational power of the ansatz, but the trainability under the relaxed objective function.
Motivated by this finding, we propose a multistage continuation framework that gradually transforms a smooth relaxed objective into a sharper objective that more closely approximates the target discrete problem.
Numerical experiments on G-set instances with 800 vertices show that the proposed method consistently outperforms conventional PCE and is competitive with representative graph neural network (GNN) methods.
These results clarify the main factors behind PCE performance and provide a practical strategy for improving PCE on quantum devices with limited qubit resources.
\end{abstract}

\maketitle
\section{Introduction}
Combinatorial optimization problems are ubiquitous in science and industry, arising in applications such as logistics, finance, network design, and resource allocation.
Because many such problems are computationally hard, a wide range of classical approaches such as local search, semidefinite programming (SDP), and simulated annealing (SA) have been developed.
In parallel, quantum optimization has attracted growing interest as advances in quantum hardware have enabled the exploration of increasingly large problem instances on near-term devices~\cite{Abbas2024Challenges,Ebadi2022MISRydberg}.

Among quantum approaches, quantum annealing (QA)~\cite{PhysRevE.58.5355,Yarkoni_2022} and variational quantum algorithms (VQAs)~\cite{vqa} such as the Quantum Approximate Optimization Algorithm (QAOA)~\cite{vqa,qaoa} have emerged as two major paradigms for combinatorial optimization.
QA generally relies on dedicated hardware that exploits quantum effects during an annealing process.
By contrast, QAOA uses parameterized quantum circuits for current quantum devices, with cost and mixer operators designed according to the target problem.
Despite this flexibility, conventional QAOA typically assigns one qubit to each binary variable.
Consequently, the required number of qubits grows linearly with the number of variables, limiting the size of the instances that can be treated on current devices.

To address this limitation, Pauli Correlation Encoding (PCE)~\cite{Sciorilli2025Towards} was recently introduced to encode binary variables in the expectation values of Pauli strings while reducing the number of required qubits.
For a fixed correlation order, the number of encoded variables can grow polynomially with the number of qubits.
Beyond reducing qubit requirements, this compressed representation has also been argued to improve optimization behavior by mitigating barren plateau effects~\cite{mcclean2018barren}.
PCE has since been extended to applications including the traveling salesman problem~\cite{carmo2025warmstartingpcetravelingsalesman}, optimization under budget constraints~\cite{pcebudget}, the LABS problem~\cite{sciorilli2026competitivenisqqubitefficientsolver}, and portfolio optimization~\cite{soloviev2025largescaleportfoliooptimizationusing}.

Although PCE has demonstrated competitive performance compared with standard classical methods such as semidefinite programming (SDP), the mechanisms underlying this performance remain unclear.
Leaving aside its implementation on quantum hardware, PCE can be viewed as a nonlinear continuous relaxation in which continuous parameters are optimized to produce relaxed representations of binary variables.
This viewpoint motivates a comparison with classical parametric models and raises the question of whether the performance of PCE can be explained by properties shared with such models or instead depends on features specific to quantum circuits, including circuit connectivity and gate constraints.

In this work, we first address this question by comparing PCE with a hierarchy of classical surrogate models based on tensor networks whose internal structures progressively approach the structure of the PCE circuit.
This comparison allows us to examine the roles of parameter count, connectivity, and constraints on local transformations.
We then investigate the gap between conventional PCE and the best-known solutions to Max-Cut.
To do so, we consider two possible sources of this gap. One is insufficient expressivity of the circuit ansatz, and the other is poor trainability under the conventional relaxed objective function. 
To distinguish between these possibilities, we conduct an expressivity test by introducing a supervised objective function that trains the circuit toward reference configurations for Max-Cut.
The results show that even shallow circuits are expressive enough to represent the best-known solutions.
This finding indicates that the dominant limitation is not the representational capacity of the ansatz but the inability of the conventional objective function to guide the optimization toward parameter regions that yield solutions with large cut values.

Based on this understanding, we focus on the gap between the discrete and relaxed objective functions in PCE optimization and propose a multistage continuation framework.
The central idea is to begin the optimization with a smoother objective function that is easier to optimize and then gradually transform it into a sharper objective function that more closely approximates the original discrete optimization problem.
This schedule improves the optimization pathway without modifying the circuit ansatz itself, thereby directly targeting the trainability bottleneck identified in our analysis.
Finally, we compare multistage PCE with conventional PCE and representative GNN methods that optimize continuous trainable parameters like PCE. 
We consider the Max-Cut problem and select several instances with 800 vertices from the G-set benchmark.
On eight instances from the G-set benchmark with 800 vertices, multistage PCE improves the cut value over conventional PCE in every case.
It also outperforms both GNN baselines on G1, G2, and G3 and attains the best known value on G1.
These results provide evidence that trainability under the relaxed objective function is a primary source of performance saturation and offer a systematic strategy for improving PCE.

The remainder of this paper is organized as follows.
In Sec.~\ref{sec:preliminary}, we review the basic formulation of PCE used throughout this work.
In Sec.~\ref{sec:TN}, we introduce the comparative framework based on tensor networks and analyze how parameter efficiency and quantum circuit topology contribute to the performance of PCE.
In Sec.~\ref{sec:expressivity}, we study the expressivity of the PCE ansatz through supervised learning toward reference Max-Cut solutions.
In Sec.~\ref{sec:kstage}, we propose a multistage continuation framework to improve trainability and evaluate its effect numerically.
In Sec.~\ref{sec:gnn}, we benchmark the resulting multistage PCE against representative GNN methods on G-set instances.
Finally, we conclude in Sec.~\ref{sec:conclusion}.

\section{Preliminaries}
\label{sec:preliminary}
\subsection{Pauli Correlation Encoding (PCE)}
Pauli Correlation Encoding (PCE)~\cite{Sciorilli2025Towards} is a framework for variational quantum optimization that reduces the number of required qubits.
Its central idea is to encode classical optimization variables into the expectation values of Pauli strings, rather than assigning one qubit to each variable.

Let $\Pi=\{\Pi_i\}_{i=1}^m$ be a set of traceless Pauli strings acting on a system of $n$ qubits, and let $[m]=\{1,\ldots,m\}$.
In PCE, the binary variable associated with $\Pi_i$ is defined as
\begin{align}
x_i \coloneqq \mathrm{sgn}(\langle \Pi_i \rangle), \qquad i \in [m],
\label{eq:pce_decode}
\end{align}
where
\begin{align}
\langle \Pi_i \rangle \coloneqq \langle \Psi | \Pi_i | \Psi \rangle
\end{align}
is the expectation value of $\Pi_i$ in the quantum state $|\Psi \rangle$.

Following the conventional PCE construction, we focus on the subset $\Pi^{(k)}$ consisting of weight-$k$ Pauli strings formed by permutations of 
\(
X^{\otimes k}\otimes I^{\otimes(n-k)},
\)
\(
Y^{\otimes k}\otimes I^{\otimes(n-k)},
\)
and
\(
Z^{\otimes k}\otimes I^{\otimes(n-k)}.
\)
This choice is experimentally convenient because $\Pi^{(k)}$ is the union of three subsets, each of which is internally commuting, allowing all encoded correlations to be estimated using only three measurement settings.

Using all possible permutations yields
\begin{align}
m = 3 \binom{n}{k} = O(n^k),
\label{eq:pce_scaling}
\end{align}
so, for fixed $k$, the number of encodable variables grows polynomially with the number of qubits.
Equivalently, a problem with $m$ binary variables can be encoded using only $ n= O(m^{1/k})$ qubits.

We consider the weighted Max-Cut problem on an undirected graph $G=(V,E)$ with edge weights $w_{ij}$.
Let $x_i \in \{+1,-1\}$ denote the spin associated with vertex $i$.
The cut value can be written as
\begin{align}
C(\boldsymbol{x}) = \frac{1}{2}\sum_{(i,j)\in E} w_{ij} (1-x_i x_j).
\label{eq:objective}
\end{align}
Maximizing $C(\boldsymbol{x})$ is therefore equivalent to minimizing the corresponding Ising interaction term.
Because the sign function is nondifferentiable, PCE replaces it with a hyperbolic tangent and minimizes the relaxed objective function
\begin{gather}
\mathcal{L}
=
\sum_{(i,j)\in E}
w_{ij}\,
\tanh(\alpha \langle \Pi_i \rangle)\,
\tanh(\alpha \langle \Pi_j \rangle)
+
\mathcal{L}^{(\mathrm{reg})},
\label{eq:total_loss}
\\
\mathcal{L}^{(\mathrm{reg})}
=
\beta \nu
\left[
\frac{1}{m}
\sum_{i=1}^{m}
\tanh^2\!\bigl(\alpha \langle \Pi_i \rangle\bigr)
\right]^2.
\label{eq:reg_term}
\end{gather}
Here, the first term is a continuous relaxation of the Ising interaction term obtained by replacing each binary spin with its relaxed PCE representation.
The second term is a regularization term that encourages the expectations of correlators to remain small during training. 
The factor $1/m$ normalizes the quantity in brackets to $O(1)$, and $\nu$ is chosen according to the problem so that the regularization term has a magnitude comparable to that of the first term.
The parameter $\beta$ controls the strength of the regularization, while $\alpha$ controls the steepness of the hyperbolic tangent. A larger $\alpha$ makes the relaxed mapping closer to the sign function, whereas a smaller $\alpha$ gives a smoother objective landscape.

Unless otherwise stated, for weighted Max-Cut we choose
\begin{align}
\nu = \frac{W}{2} + \frac{W_{\mathrm{MST}}}{4},
\end{align}
where $W=\sum_{(i,j)\in E} w_{ij}$ is the total edge weight and $W_{\mathrm{MST}}$ is the weight of a minimum spanning tree of the graph ~\cite{poljak1986heuristic}.
For unweighted Max-Cut, this reduces to the Edwards--Erd\H{o}s bound~\cite{bylka1999maximum}
\begin{align}
\nu = \frac{|E|}{2} + \frac{|V|-1}{4}.
\end{align}
For $\alpha$ and $\beta$, we follow the conventional empirical setting $\alpha\approx n^{k/2}$ and $\beta=0.5$.

In practice, a parameterized quantum circuit prepares the state $|\Psi(\boldsymbol{\theta})\rangle$.
Typically, fixed circuit ansatzes such as a brick-wall ansatz and an all-to-all entangling ansatz are used.
The circuit consists of repeated applications of a prescribed layer pattern, and we denote by $p$ the number of repeated layers.
After training, the binary variables are decoded using Eq.~\eqref{eq:pce_decode}.
The resulting solution may optionally be refined through classical local search.

\section{Structural Analysis via Tensor-Network Models}
\label{sec:TN}
In this section, we investigate how the internal structure of PCE contributes to its performance by viewing PCE as a parametric model.
From this perspective, the parametrized quantum circuit defines a structured mapping from trainable parameters to Pauli expectation values, which are used to construct the nonlinear relaxed objective function.
This view allows us to place PCE within a broader class of continuous optimization frameworks and permits comparison with classical surrogate models.
Specifically, we construct a hierarchy of tensor network (TN) models whose structures progressively approach the topology of the PCE circuit.
By comparing these models, we systematically examine how parameter count and circuit connectivity contribute to the performance of PCE.

\subsection{Construction of Comparative Tensor Network Models}
% \if 0
% Building on this perspective, we introduce an analytical framework utilizing Tensor Networks (TNs) to represent the linear transformation layers of the NN, aiming to elucidate the structural origins of the performance of PCE.
% TNs, which originated in the study of quantum many-body systems ~\cite{orus2014practical}, are highly effective for the compressed representation of high-dimensional linear operators.
% They provide a natural tool for precisely emulating the specific connectivity patterns of quantum circuits without sacrificing the expressive power of classical neural networks.
% In this section, starting from a classical fully connected NN, we construct three comparative models by progressively transitioning their internal structures toward those of a quantum circuit (see Fig.~\ref{fig:tn_models}).
% It is important to note that this theoretical framework is inherently universal.
% The structural transition from a dense fully connected network to a localized quantum circuit is entirely continuous.
% Because complex global connections are naturally present in a fully connected network, this framework allows us to interpret the quantum circuit as a sparsified version of a dense network subject to locality constraints.
% Through this approach, we structurally isolate and verify the factors underpinning the performance of PCE.
% \fi

\begin{figure}[tbp]
    \centering
    \hspace{5cm}
    \includegraphics[width=0.9\linewidth]{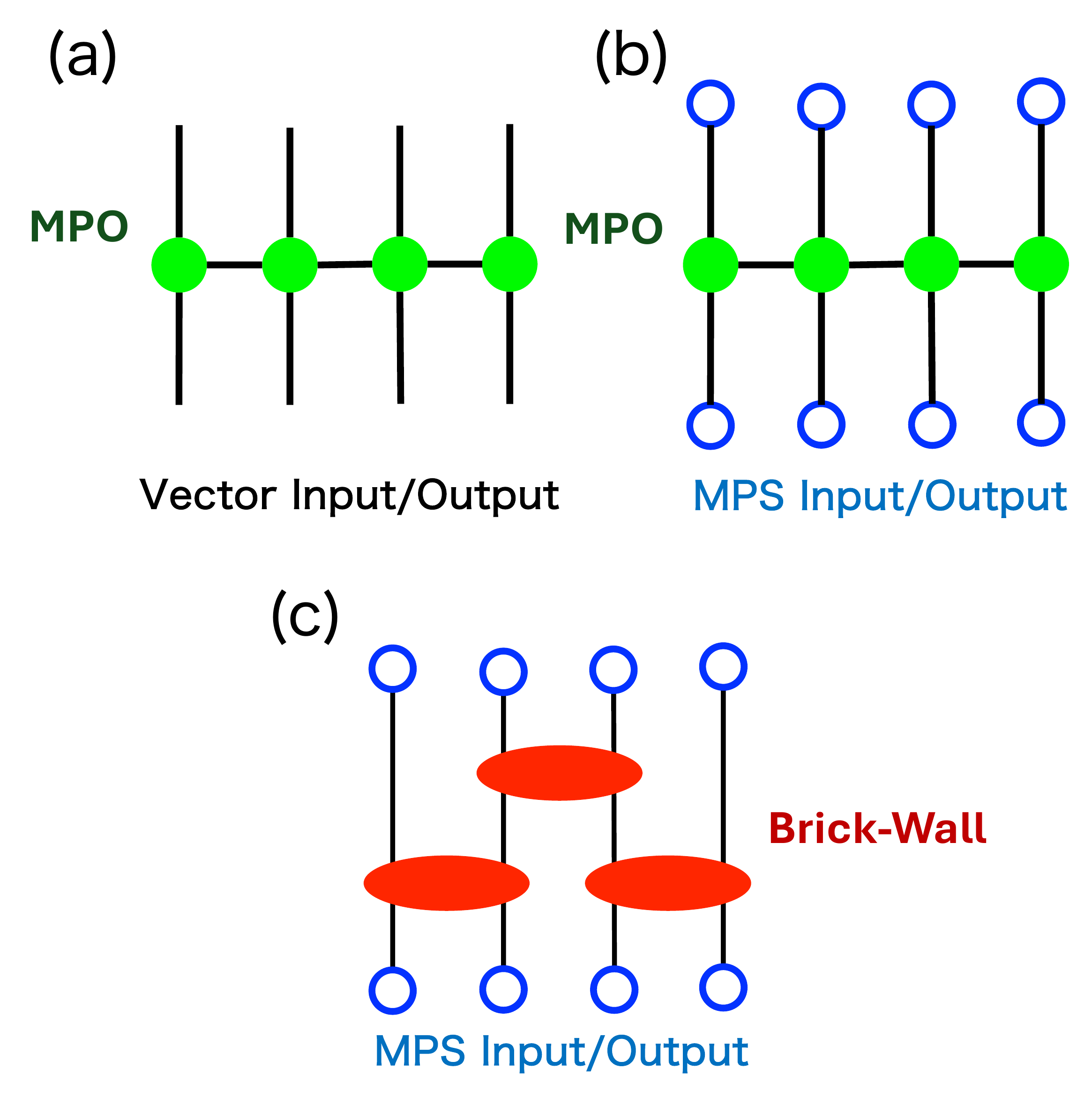} 
    \caption{Three comparative models based on tensor networks introduced in this study. (a) MPO-based model, (b) MPS-based model, and (c) Brick-wall model. These models are designed so that their internal structures progressively approach the topology of the PCE circuit.}
    \label{fig:tn_models}
\end{figure}

To identify the structural factors underlying the performance of PCE optimization, we introduce a unified framework in which the linear transformation layers are represented using tensor networks (TNs)~\cite{orus2014practical}.
A dense $m\times m$ weight matrix provides a natural reference model but requires $O(m^2)$ trainable parameters, which is far larger than the parameter budgets used for PCE in this work.
We therefore use TN decompositions to construct compressed classical models with substantially fewer parameters.
Within this setting, in which the parameter counts are controlled, we progressively impose structures that more closely resemble a quantum circuit, as illustrated in Fig.~\ref{fig:tn_models}.
The three models are (a) an MPO model with classical vector inputs and outputs, (b) an MPS model with a tensor network state representation and outputs based on Pauli expectation values, and (c) a brick wall TN model whose connectivity matches that of the PCE circuit.

\paragraph{MPO-based Model}
The primary barrier when comparing a standard fully connected NN with PCE is the difference in parameter scaling.
A dense $m\times m$ weight matrix requires $O(m^2)$ trainable parameters. 
We therefore use an MPO decomposition to reduce the parameter count to a range comparable to that of PCE. 
An MPO decomposes a large matrix into a chain of local tensors. 
When the local dimensions and bond dimension are fixed, the number of parameters can scale linearly with the system size.
By employing this decomposition, we restrict the parameter count to match the scale of PCE.
The inputs and outputs are treated as standard classical vectors.
Let $\boldsymbol{x} = (x_1, x_2, \dots, x_m)$ be the continuous output vector of the MPO-based model, where each element $x_i \in [-1, 1]$ directly corresponds to the Pauli expectation value $\langle \Pi_i \rangle$ used in the PCE framework.
To optimize this model, we employ a loss function $\mathcal{L}_{\mathrm{MPO}}$, which is a simplified version of the conventional PCE objective function with the regularization term removed:
\begin{equation}
    \mathcal{L}_{\mathrm{MPO}} = \sum_{(i,j) \in E} w_{ij} x_i x_j,
    \label{eq:MPO}
\end{equation}
where $E$ represents the set of edges and $w_{ij}$ denotes the weight of the edge between nodes $i$ and $j$.
This continuous relaxation allows for gradient-based optimization while serving as a baseline to compare classical NNs and PCE from the standpoint of parameter efficiency.
\paragraph{MPS-based Model}
To make the model closer to the quantum formulation, we represent the inputs and outputs as matrix product states rather than ordinary vectors.
An MPS represents a state vector with many components as a sequence of local tensors and can capture correlations among local degrees of freedom.
The operator is represented by a generic MPO without constraints on its connectivity.
For the loss function, we use Eq.~\eqref{eq:MPO}, as in PCE, with the Pauli expectation values evaluated from the output MPS.
This model, therefore, examines the combined effect of the MPS representation and the PCE objective function.
\paragraph{Brick-wall Model}
To make the model structure closer to that of the actual quantum circuit, we replace the generic MPO connectivity with the brick wall tensor network structure used in PCE.
However, the local tensors corresponding to gates remain generic complex matrices and are not constrained to be unitary or to follow the specific parameterized rotation structure of PCE.
Comparing this model with PCE allows us to examine the combined effects of circuit topology, unitary constraints, and gate parameterization.

\subsection{Numerical Simulations and Structural Analysis}
\begin{figure}[tbp]
    \centering
    \hspace{-2cm}
    \includegraphics[width=0.9\linewidth]{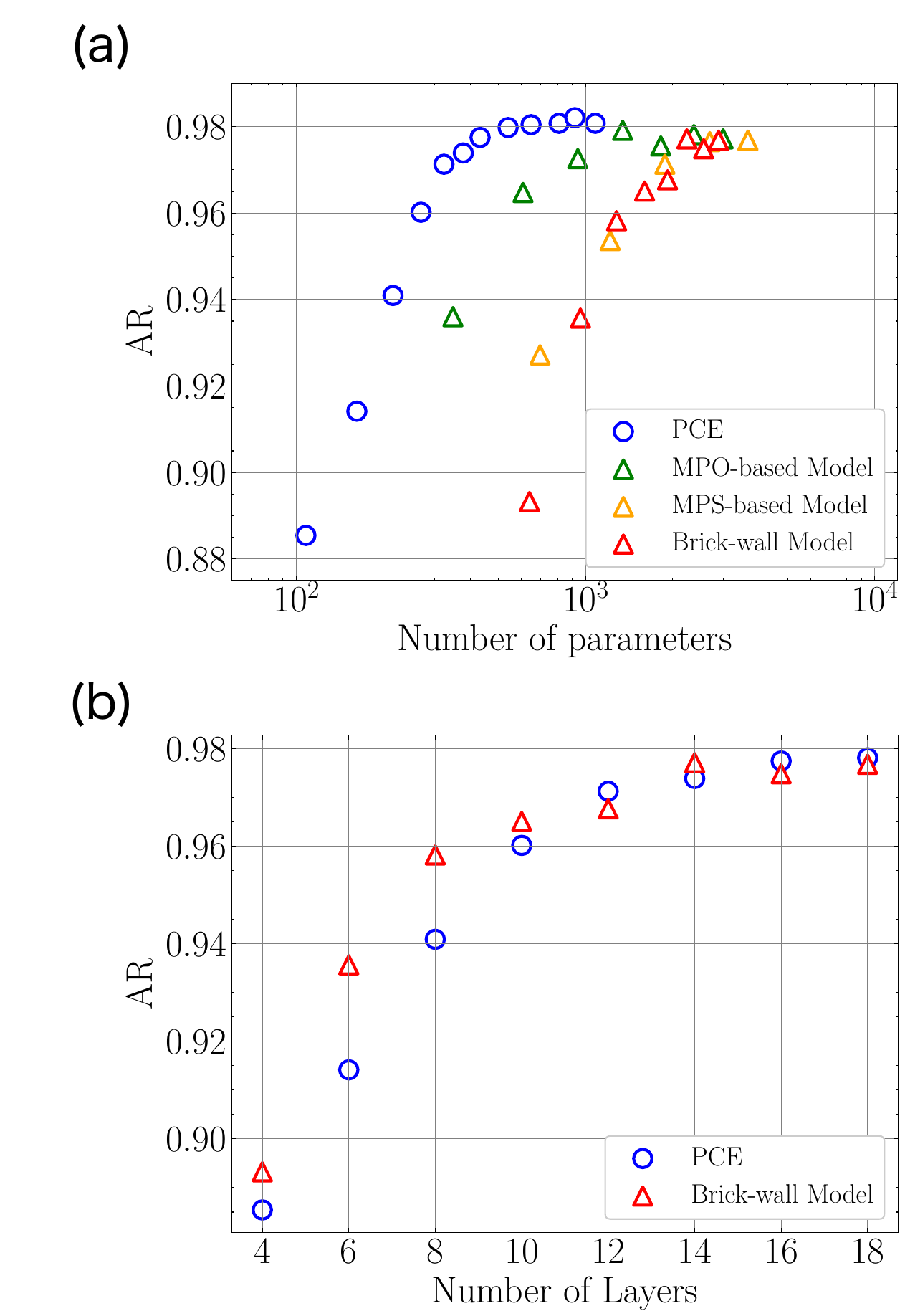} 
    \caption{Performance evaluation of the comparative models on the G14 benchmark instance ($|V|=800$).
    (a) Parameter Efficiency: The relationship between the total number of parameters (log scale) and the achieved approximation ratio across all models.
    (b) Structural Comparison: The approximation ratio as a function of the number of layers, comparing the PCE Brick-wall ansatz and the Brick-wall model based on tensor networks, both sharing the same brick-wall connectivity.}
    \label{fig:result_tn}
\end{figure}

We now analyze the simulation results from the perspectives of parameter efficiency and circuit topology.
In the following simulations, we use a PCE setting with $n=11$ qubits and $k=6$.
For the structural comparison, we adopt the brick-wall ansatz on the PCE side and compare it with Tensor Network models constructed with the corresponding layered structure.
As a benchmark problem, we consider the Max-Cut problem on G14 ($|V|=800$), an instance from the G-set~\cite{Gset}, a widely used benchmark set for Max-Cut.
The performance of each model is evaluated in terms of the approximation ratio (AR), computed relative to the best-known solution reported in the literature~\cite{benlic2013breakout}.
Each simulation is repeated five times with independent random initializations, and we report the maximum AR obtained across the five runs.
The results are summarized in Fig.~\ref{fig:result_tn}.

Figure~\ref{fig:result_tn}(a) shows the AR as a function of the total number of trainable parameters for PCE and the comparative TN models (MPO-based, MPS-based, and Brick-wall models).
While all models generally improve as parameter count increases and reach similar AR values around $0.98$, a clear difference in parameter efficiency is observed.
Among the TN models, the MPS-based and Brick-wall models, which represent inputs and outputs as quantum states, require significantly more parameters than the simpler MPO-based model to achieve the same AR, even though the MPO-based model treats them as classical vectors.
This suggests that using generic complex tensors to represent quantum states can unnecessarily enlarge the search space unless suitable structural restrictions, such as circuit locality and unitary gate constraints, are imposed.
The comparison between the brick-wall model and PCE allows us to examine the combined effects of unitary constraints and gate parameterization.
Thus, simply forcing a model into a quantum-like TN format without careful parameterization can lead to an inefficient allocation of trainable parameters.

In contrast, PCE achieves comparable or superior performance with a substantially smaller parameter count.
This result highlights the importance of parameter allocation.
The Brick-wall model based on tensor networks shows that using additional parameters to make each local tensor more expressive yields only limited performance gains.
By contrast, PCE restricts local operators to simple parameterized rotation gates and can allocate the saved parameter budget to increasing the number of layers.
These results suggest that search performance is influenced more strongly by the propagation of correlations through circuit depth and topology than by the local expressivity of individual gates.

To further examine the role of circuit topology and correlation propagation, we directly compare the PCE brick-wall ansatz with the brick-wall model based on tensor networks. 
Both are constructed with the same brick-wall connectivity.
Figure~\ref{fig:result_tn}(b) plots the approximation ratio against the number of layers for both models.
The two performance curves show similar behavior.
Because the Brick-wall model based on tensor networks consists of generic matrices without unitary constraints, it possesses approximately five times as many parameters per gate as PCE, giving it greater local flexibility.
Indeed, at very shallow depths, this expressive advantage allows the Brick-wall model based on tensor networks to achieve a slightly higher approximation ratio.
However, as the number of layers increases, this gap rapidly narrows, and both models reach similar approximation ratios.

This convergence strongly suggests that high local expressivity is not the dominant factor governing search performance.
If local expressivity were the primary driver, the highly flexible Brick-wall model based on tensor networks would consistently outperform PCE.
% Instead, the fact that equivalent depths yield equivalent solutions under the same topological structure leads to the definitive conclusion that the most critical role in the solution search is played by the entanglement generated through the brick-wall structure.
Instead, the fact that equivalent depths yield comparable solutions under the same topological structure suggests that circuit topology and the resulting propagation of correlations play a central role in the search process.

The relationship between the circuit structure and the PCE loss function provides further theoretical support for this conclusion.
Generally, an unconstrained operation on an $N$-qubit state is represented by a $2^N \times 2^N$ matrix.
% Even a fully parameterized two-qubit operation requires up to 16 parameters.
By restricting interactions to local gates, PCE reduces the number of trainable parameters while retaining the ability to control Pauli expectation values across multiple layers.
Because PCE represents binary variables using expectation values of multi-qubit Pauli strings, its performance depends on how effectively the ansatz can control these expectation values. 
The results above suggest that this control is achieved not by increasing the local expressivity of individual gates, but by using simple parameterized gates over multiple layers to propagate correlations across qubits in a parameter-efficient manner.
Thus, the performance of PCE can be understood as resulting from an effective allocation of trainable parameters between local gate expressivity and circuit depth.

\section{Expressivity Analysis via Supervised Learning}
\label{sec:expressivity}

Conventional PCE has been shown to achieve solution quality comparable to that of efficient classical optimization methods. However, a gap between its performance and the best-known solutions remains.
A remaining question in understanding the limitations of conventional PCE is whether its performance is constrained by the representational power of the ansatz, namely expressivity, or by the difficulty of the optimization process, namely trainability.
To separate these factors, we conduct an expressivity test by introducing a supervised objective function to train the circuit toward reference Max-Cut solutions.
If the supervised objective function can be reduced to close to zero with a shallow circuit, this indicates that the ansatz is sufficiently expressive to represent strong solutions.
In that case, the remaining limitation of conventional PCE is more likely due to poor trainability under the unsupervised relaxed objective function rather than insufficient expressivity.

\subsection{Formulation of the Supervised Loss}

Let $\mathbf{x}^*\in\{-1,+1\}^m$ denote a target spin configuration associated with the best-known solution for a given instance.
We define the supervised loss function as the mean squared error between the target and the relaxed PCE outputs:
\begin{align}
\mathcal{L}_{\mathrm{sup}}(\boldsymbol{\theta})
=
\frac{1}{m}\sum_{i=1}^{m}
\left(
x_i^*
-
\tanh\!\bigl(\alpha \langle \Pi_i \rangle \bigr)
\right)^2,
\label{eq:supervised_loss}
\end{align}
where $\langle \Pi_i \rangle = \langle \Psi(\boldsymbol{\theta}) | \Pi_i | \Psi(\boldsymbol{\theta}) \rangle$.
Minimizing $\mathcal{L}_{\mathrm{sup}}$ explicitly tests whether the chosen ansatz contains parameters that approximately reproduce the target configuration.
The supervised optimization is not intended as a practical solver because the target solution is generally unknown. Instead, it is used only to assess the expressivity of the ansatz.

\subsection{Experimental Verification}
\begin{figure}[t]
\centering
\includegraphics[width=\linewidth]{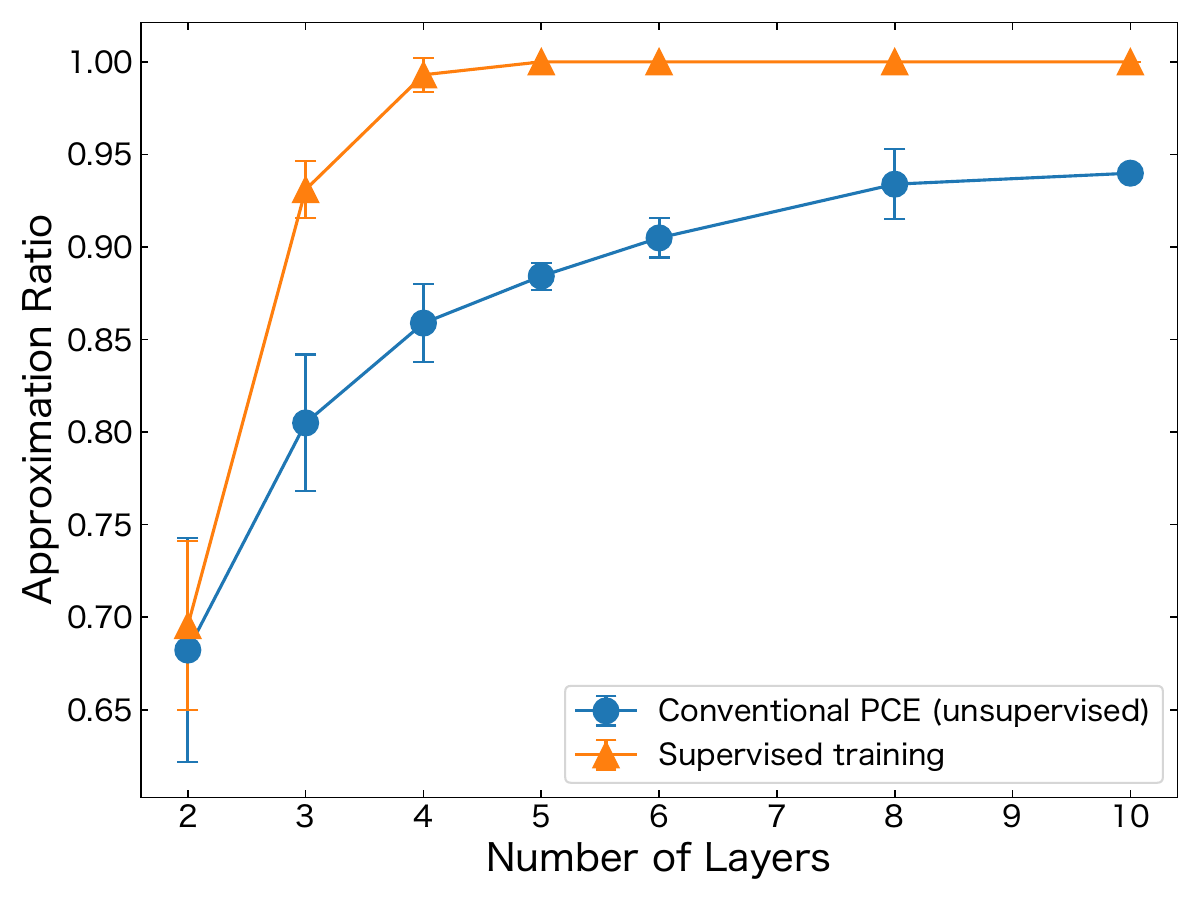}
\caption{Approximation ratio as a function of circuit depth $p$ for conventional unsupervised PCE and supervised optimization.
Each point represents the approximation ratio averaged over three G-set instances with 800 vertices, G6, G11, and G18.
For each instance, the best result obtained over five random initializations is used.
Error bars indicate the standard deviation across the three instances.}
\label{fig:teaching}
\end{figure}

We compare the unsupervised optimization of Eq.~\eqref{eq:total_loss} with the supervised optimization of Eq.~\eqref{eq:supervised_loss} on instances with 800 vertices from the G-set benchmark.
Throughout this section, we use the all-to-all ansatz with $n=13$ qubits and $k=3$. 
For the supervised optimization, we use a reference spin configuration corresponding to the best-known solution for each instance reported in Ref.~\cite{benlic2013breakout}.

Figure~\ref{fig:teaching} shows that supervised optimization reaches $\mathrm{AR}=1.0$ with only five layers, whereas unsupervised optimization saturates at a lower AR despite using the same ansatz.
This indicates that the ansatz is expressive enough to represent strong solutions, and that the main limitation in the conventional setting is not insufficient expressivity but poor trainability under the relaxed unsupervised objective function.
This further suggests that performance gains may be possible if one can design unsupervised objective functions that induce optimization dynamics similar to those of the supervised objective function.

\begin{figure}[t]
\centering
\includegraphics[width=\linewidth]{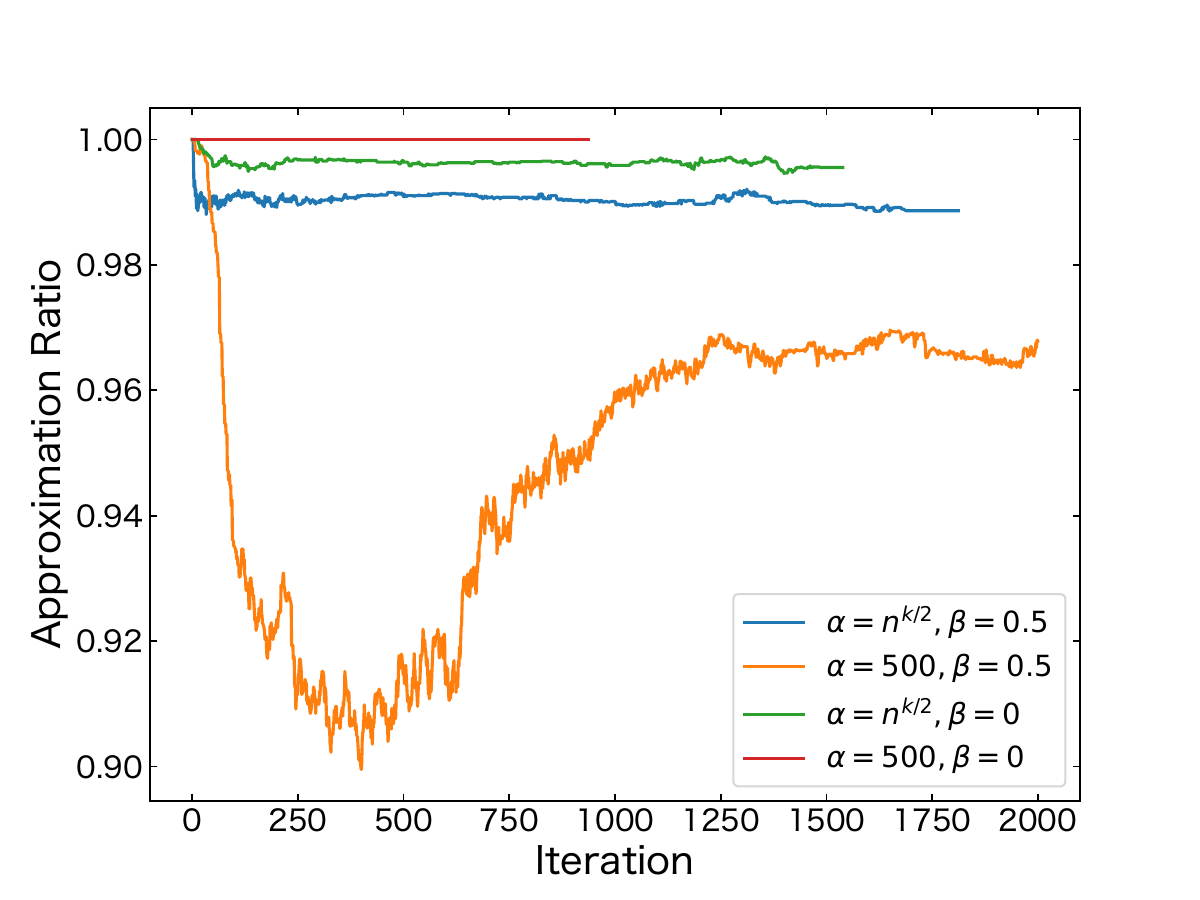}
\caption{Optimization trajectories starting from circuit parameters that reproduce the best-known solution (AR$=1$ at iteration $0$). Different choices of $(\alpha,\beta)$ lead to markedly different stability near this solution-reproducing region on the G1 instance.}
\label{fig:teacher_init}
\end{figure}

To see this more clearly, Fig.~\ref{fig:teacher_init} shows that even when the optimization is started from parameters that provide the best-known solution, the trajectory under the unsupervised relaxed objective function deviates from AR$=1$ depending on the choice of the hyperparameters $(\alpha,\beta)$. 
In particular, the conventional setting $(\alpha,\beta)=(n^{k/2},0.5)$ leads to a decrease in AR from $1$, whereas $(n^{k/2},0)$ and $(500,0)$ mitigate this effect. 
This trend is consistent with the fact that increasing $\alpha$ and decreasing $\beta$ bring the relaxed objective function in Eq.~\eqref{eq:total_loss} closer to the discrete objective function in Eq.~\eqref{eq:objective}.
As a result, the optimization remains near parameter regions that yield strong solutions.
These observations indicate that the main limitation is not the absence of parameters corresponding to strong solutions, but rather the inability of the unsupervised objective to guide the optimization toward such parameters.
This directly motivates the next section, where we modify the optimization process itself through stage-wise scheduling of the objective function.

\section{Improving Trainability through Multistage Hyperparameter Scheduling}
\label{sec:kstage}
The expressivity analysis in Sec.~\ref{sec:expressivity} suggests that the main limitation of conventional PCE is not insufficient circuit expressivity, but rather the difficulty of optimization under the  unsupervised objective function.
To address this issue, we introduce a multistage continuation framework, in which optimization begins with a smoother objective function and gradually transitions to a sharper one that better reflects the target discrete solution.
In the following, we formulate this idea as a general optimization framework with $K$ stages for PCE.

\subsection{General Formulation with \texorpdfstring{$K$}{K} Stages}
Let $\boldsymbol{\theta}$ denote the circuit parameters, and let $\mathcal{L}(\boldsymbol{\theta};\alpha,\beta)$ denote the PCE objective function in Eq.~\eqref{eq:total_loss}, where $\alpha$ controls the steepness of the tanh activation and $\beta$ controls the strength of the regularization term.
In the formal limit $\alpha \to \infty$ and $\beta \to 0$, the relaxed objective approaches the discrete Ising objective that is equivalent to maximizing the Max Cut value in Eq.~\eqref{eq:objective}, which motivates multistage hyperparameter scheduling.

At each stage $s \in \{1,2,\dots,K\}$, we optimize the objective function for stage $s$
% \begin{align}
%     \boldsymbol{\theta}^{(s)}
%     =
%     \arg\min_{\boldsymbol{\theta}}
%     \mathcal{L}(\boldsymbol{\theta};\alpha_s,\beta_s),
%     \qquad
%     \boldsymbol{\theta}^{(s)}_0 = \boldsymbol{\theta}^{(s-1)},
% \end
\begin{align}
    \boldsymbol{\theta}^{(s)}
    \approx
    \arg\min_{\boldsymbol{\theta}}
    \mathcal{L}(\boldsymbol{\theta};\alpha_s,\beta_s),
\end{align}
using the optimized parameters from the previous stage, $\boldsymbol{\theta}^{(s-1)}$, as the initial parameter vector.
The initial vector $\boldsymbol{\theta}^{(0)}$ is chosen randomly.

Accordingly, we employ a monotonically increasing schedule for $\alpha_s$ and a decreasing schedule for $\beta_s$:
\begin{align}
    \alpha_1 < \alpha_2 < \cdots < \alpha_K, \\
    \beta_1 > \beta_2 > \cdots > \beta_K.
\end{align}

The early stages use a smoother objective function to enable more stable optimization, whereas the later stages use an objective function closer to the target discrete problem.
At each stage, the parameters are optimized until the optimization has converged according to a prescribed stopping criterion. Once stage \(s\) converges, the resulting parameters \(\boldsymbol{\theta}^{(s)}\) are used to initialize stage \(s+1\).
This formulation includes simpler optimization schemes as special cases: $K=1$ corresponds to the conventional PCE optimization.

\subsection{Numerical Simulations}
\begin{figure}[tbp]
    \centering
    \includegraphics[width=\linewidth]{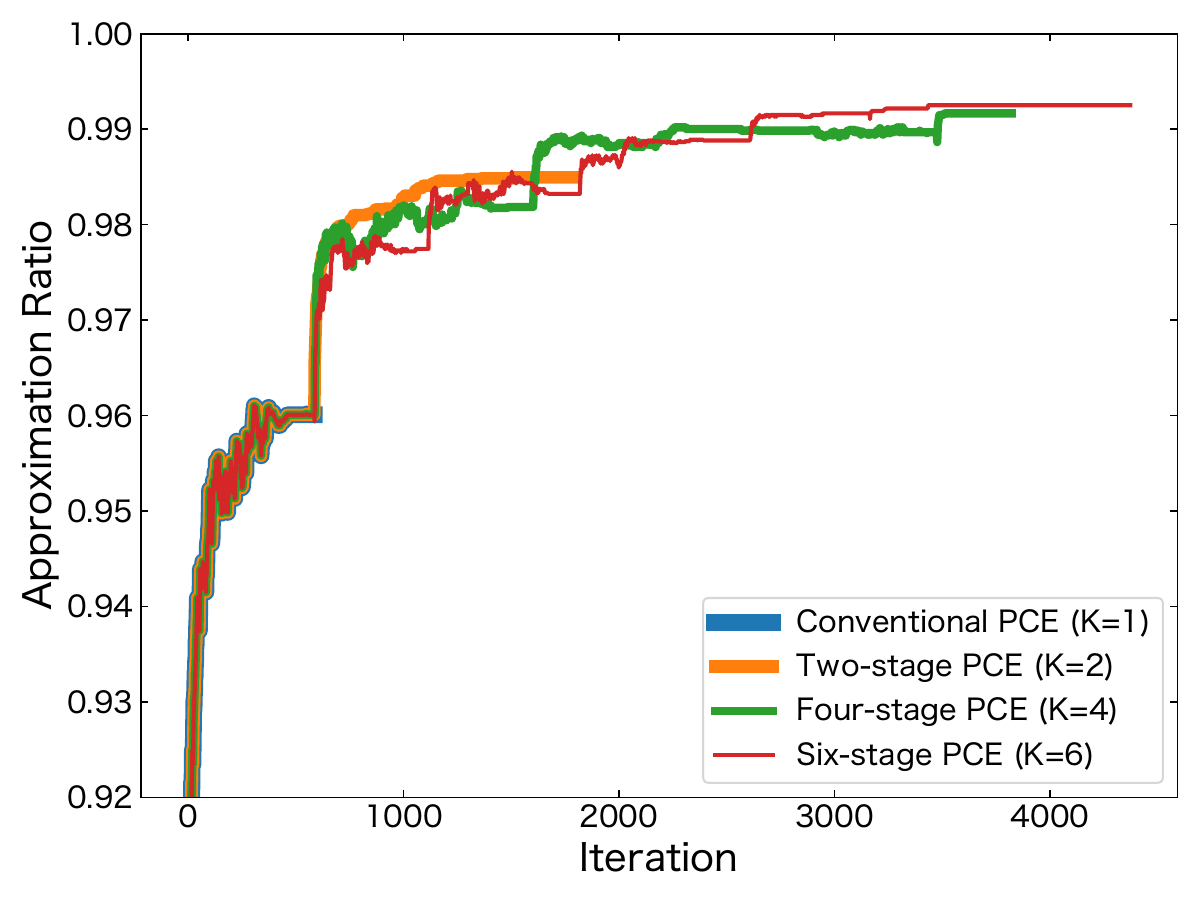}
\caption{Approximation ratio as a function of iteration for conventional PCE ($K=1$) and multistage PCE with $K=2,4,6$ on the G1 instance, using the all-to-all ansatz with fixed circuit depth $p=3$. 
All methods use the same initial circuit parameters, so their trajectories overlap in the early iterations. 
In the multistage settings, the hyperparameters are updated stage by stage using a linear schedule from $(\alpha,\beta)=(n^{k/2},0.5)$ to $(500,0)$.}
    \label{fig:trajectory}
\end{figure}

\begin{figure}[tbp]
    \centering
    \includegraphics[width=\linewidth]{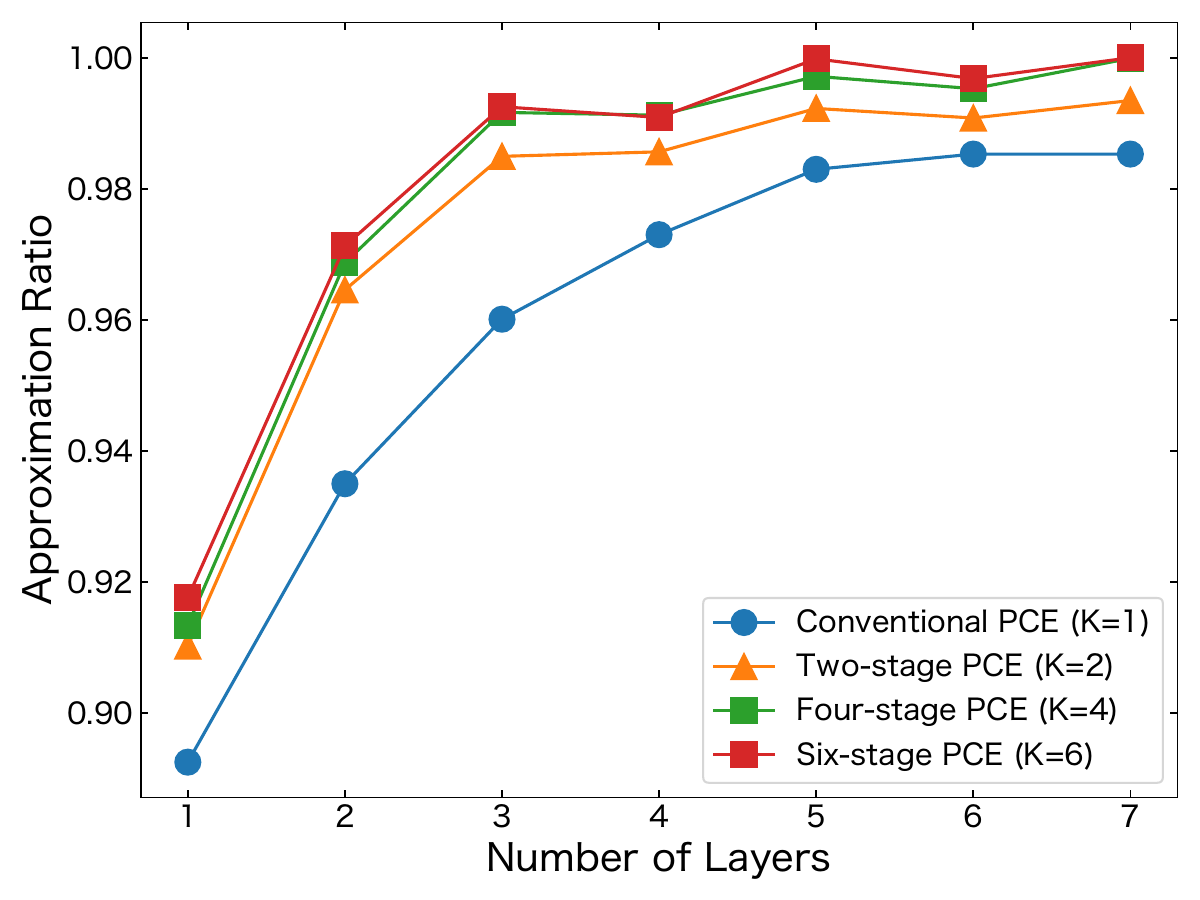}
    \caption{Approximation ratio as a function of the number of layers for conventional PCE ($K=1$) and multistage PCE with $K=2,4,6$ on the G1 instance, using the all-to-all ansatz. Each point represents the final approximation ratio obtained for the corresponding depth. All other settings are the same as in Fig.~\ref{fig:trajectory}.}
    \label{fig:k2_comparison}
\end{figure}

\begin{figure}[tbp]
    \centering
    \subfloat[End of Stage 1\label{fig:hist_step1}]{%
        \includegraphics[width=0.48\linewidth]{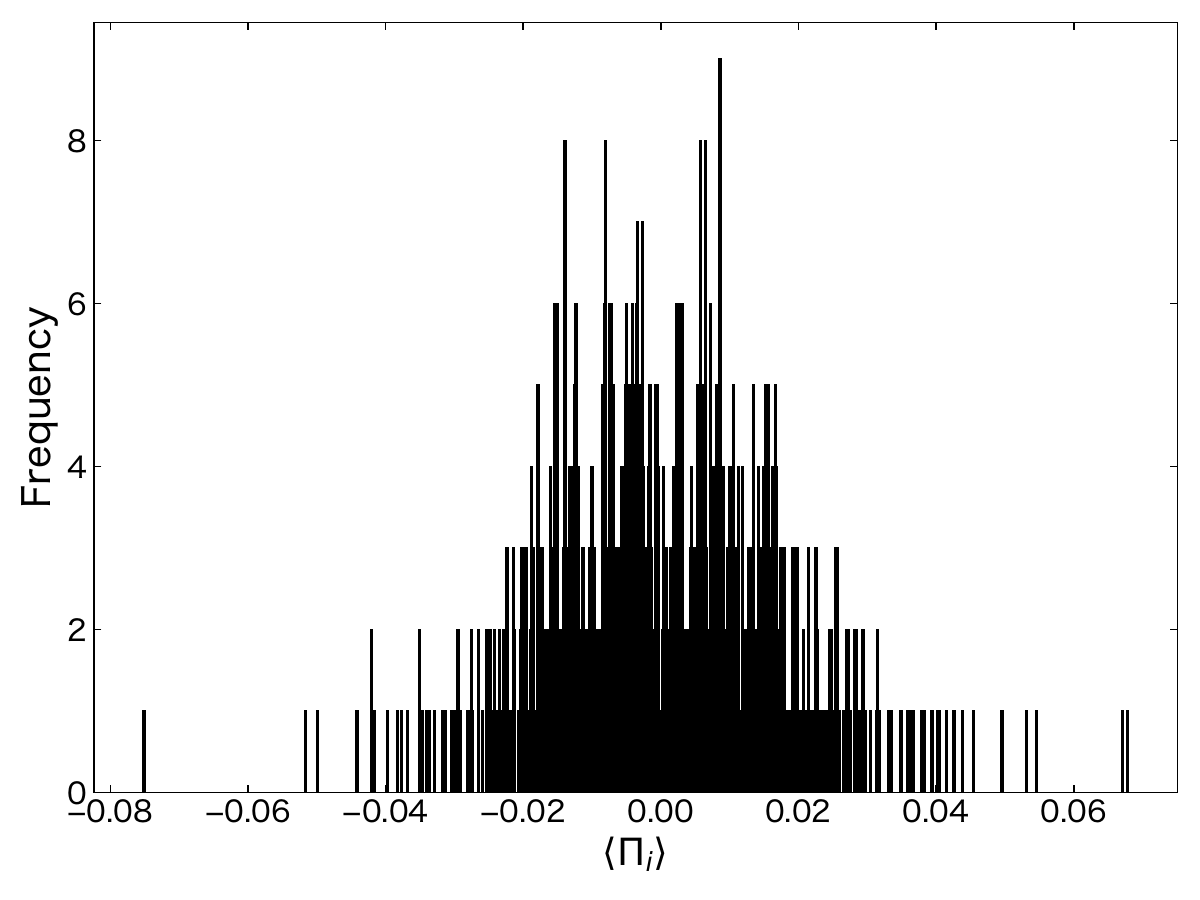}%
    }
    \hfill
    \subfloat[End of Stage 2\label{fig:hist_step2}]{%
        \includegraphics[width=0.48\linewidth]{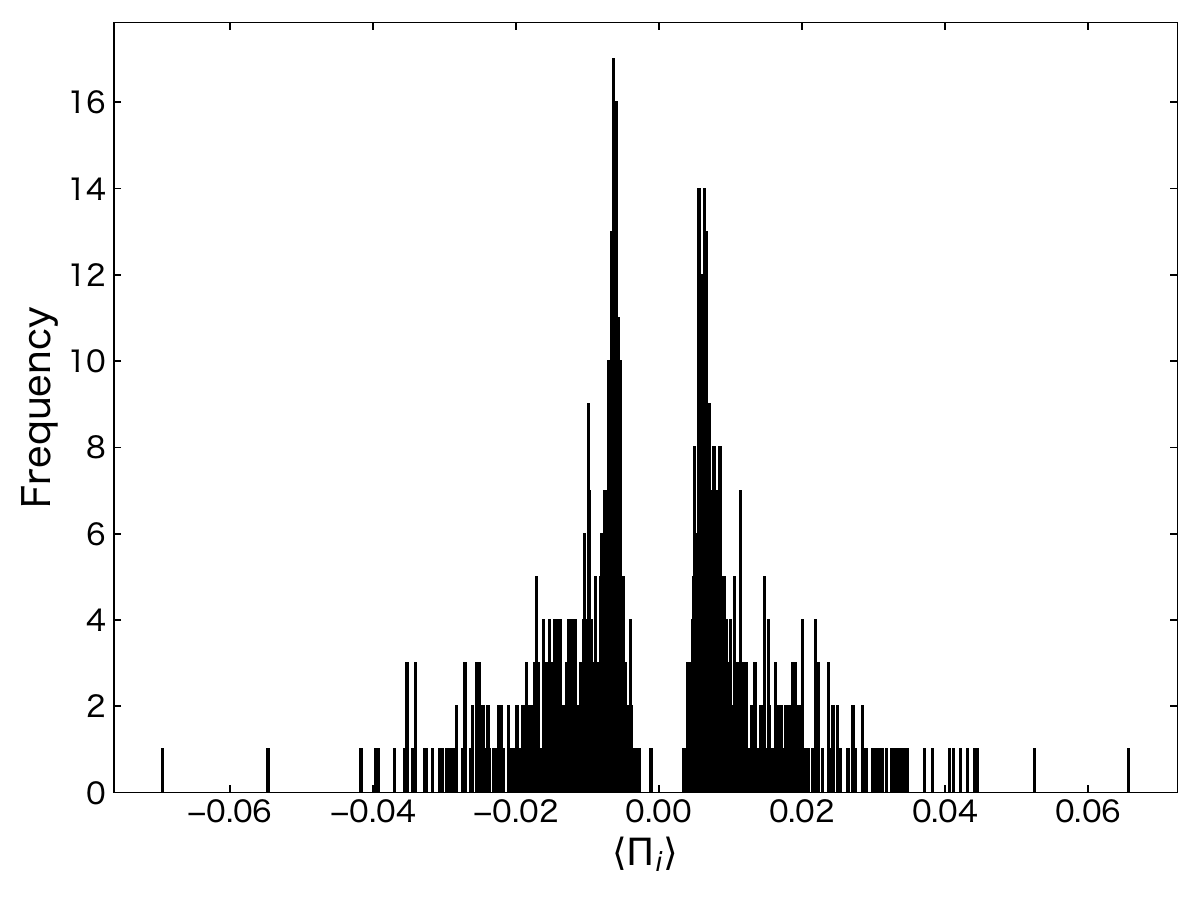}%
    }
    \caption{Histograms of the encoded variables in the case
    with two stages.}
    \label{fig:histograms}
\end{figure}
We consider the $K=2,4,6$ cases as representative examples of the proposed $K$-stage continuation framework and compare them with the conventional $K=1$ setting on the G1 instance from the G-set benchmark.
Unless otherwise stated, we use the all-to-all ansatz with $n=13$ qubits and $k=3$ in this section.

Figure~\ref{fig:trajectory} shows the approximation-ratio trajectories on the G1 instance. Because all methods start from the same initial parameters and share the same first-stage setting, their trajectories coincide in the early iterations. After the first-stage optimization has converged, however, the multistage methods change the hyperparameters and resume the optimization under a modified objective function, which leads to a further increase in AR. This behavior indicates that the change in the objective function between stages plays a significant role in improving the final solution quality.
% In this case, the optimization consists of one relaxed stage followed by one sharper stage:
% \begin{align}
%     \boldsymbol{\theta}^{(1)}
%     &=
%     \arg\min_{\boldsymbol{\theta}}
%     \mathcal{L}(\boldsymbol{\theta};\alpha_1,\beta_1), \\
%     \boldsymbol{\theta}^{(2)}
%     &=
%     \arg\min_{\boldsymbol{\theta}}
%     \mathcal{L}(\boldsymbol{\theta};\alpha_2,\beta_2),
%     \qquad
%     \boldsymbol{\theta}^{(2)}_0 = \boldsymbol{\theta}^{(1)},
% \end{align}
% with $(\alpha_1,\beta_1) = (n^{\frac{k}{2}}, 0.5)$ and $(\alpha_2,\beta_2) = (500, 0)$.
% The first-stage setting follows the conventional PCE formulation, while the second-stage setting is chosen empirically to promote stronger discretization during the final refinement stage.

Figure~\ref{fig:k2_comparison} plots the approximation ratio as a function of the number of layers on the G1 instance. 
All multistage settings ($K=2,4,6$) outperform the conventional $K=1$ optimization across the tested layer range. 
Within the optimization protocol used here, increasing the number of stages generally improves performance, and this effect is especially strong for shallow circuits. 
This is practically important because near-term quantum devices can only support limited circuit depth.
These results suggest that the bottleneck identified in Sec.~\ref{sec:expressivity} can be alleviated by improving the optimization process, even without changing the circuit ansatz itself.

The histograms for the case with two stages in Fig.~\ref{fig:histograms} provide a qualitative view of the stage-wise mechanism. 
At the end of Stage 1, the expectation values \(\langle \Pi_i \rangle\) remain in an intermediate regime rather than saturating near \(\pm 1\). 
As a result, the corresponding \(\tanh(\alpha \langle \Pi_i \rangle)\) retains sufficiently large gradients, which allows the optimization to continue making progress. 
At the end of Stage 2, the distribution of \(\langle \Pi_i \rangle\) becomes more polarized, indicating that the relaxed variables are driven closer to the discrete solution. 
This behavior is consistent with the intended transition from relaxed exploration to sharper refinement.

Overall, these observations show that substantial performance gains can be obtained without changing the circuit ansatz, by modifying the hyperparameters stage by stage and thereby reshaping the objective function during optimization. 
Combined with the supervised analysis in Sec.~\ref{sec:expressivity}, this supports the conclusion that improving trainability is a central route to enhancing PCE. 
At the same time, the stage lengths and hyperparameter schedules remain heuristic, and a more principled schedule design is an important direction for future work.

\section{Comparison with Graph Neural Networks}
\label{sec:gnn}
Through the analyses presented so far, we have identified the key factors governing the performance of PCE-based quantum optimization, particularly those responsible for its limitations, and shown that its performance can be substantially improved based on these insights.
To evaluate the practical utility of the proposed framework, we compare the resulting improved PCE method with representative graph neural network methods for graph optimization.

\subsection{Comparison with GNN methods}

\begin{table*}[t]
\centering
\caption{
Comparison of Max-Cut values on G-set benchmark instances with 800 vertices. For each method, we report the best result over 10 runs with different random seeds. The best-known values are used as the reference.
}
\label{tab:gnn_comparison}
\begin{tabular}{lccccccc}
Instance &$V$ & $E$ &  best-known & PI-GNN & QRF-GNN & Conventional PCE (K=1) & multistage PCE (K=6) \\
\hline
G1    & 800 & 19176 & 11624 & 11469 & 11609 &11480& 11624  \\
G2    & 800 & 19176 & 11620 & 11481 & 11604 &11422& 11608  \\
G3     & 800 & 19176 & 11622 & 11445 & 11617 &11420& 11621 \\
G11  & 800 & 1600  & 564   & 500 & 556   &524& 532    \\
G12  &800 & 1600  & 556   & 494 & 548   &528& 540    \\
G14    & 800 & 4694  & 3064  & 2999 & 3056  &2995& 3035   \\
G15    & 800 & 4661  & 3050  & 2972 & 3040&2996  & 3021   \\
G16    & 800 & 4672  & 3052  & 2983 & 3043  &2974& 3025   \\
\hline
\end{tabular}
\end{table*}
As classical neural-network-based baselines, we consider the Physics-Inspired GNN (PI-GNN) \cite{schuetz2022combinatorial} and its recurrent extension, QRF-GNN \cite{pugacheva2024enhancing}.
PI-GNN formulates Max-Cut as a differentiable energy minimization and optimizes an Ising-type objective in an unsupervised manner.
QRF-GNN extends PI-GNN by introducing recurrent feature updates, providing a stronger GNN method.
On the quantum side, we compare the conventional PCE ($K=1$) with the proposed multistage PCE ($K=6$).
As an absolute reference for solution quality, we use the best-known values reported by Breakout Local Search (BLS) \cite{benlic2013breakout}.
Following common practice in Max-Cut benchmarking, we evaluate how closely each learning-based method approaches these reference values.
In the experiments, we use the same number of qubits \(n\), correlation order \(k\), and circuit ansatz as in Sec.~\ref{sec:kstage}, and we set the number of stages to \(K=6\) for the multistage PCE.
Detailed hyperparameter settings and stopping criteria are provided in Appendix~\ref{app:implementation_details}.

\subsection{Simulation Results}
Table~\ref{tab:gnn_comparison} summarizes the Max-Cut values obtained on G-set benchmark instances with 800 vertices.
We first compare conventional PCE ($K=1$) with the proposed multistage PCE ($K=6$).
As shown in Table~\ref{tab:gnn_comparison}, the multistage method consistently outperforms the conventional PCE baseline on all tested instances.
The improvement is particularly pronounced on the dense instances G1--G3, where multistage PCE nearly reaches or attains the best-known values.
This result supports the conclusion of Sec.~\ref{sec:expressivity} and Sec.~\ref{sec:kstage} that improving the optimization pathway, rather than modifying the circuit ansatz itself, can substantially enhance the performance of PCE.

We next compare multistage PCE with the classical GNN methods.
Multistage PCE attains higher cut values than PI-GNN on all tested instances, despite the fact that PI-GNN explicitly exploits graph connectivity as input, whereas the PCE ansatz is not designed using the graph structure.
This indicates that the correlations generated by the quantum circuit are sufficient to achieve competitive solution quality even without an ansatz that explicitly uses graph structure.
Compared with the stronger baseline QRF-GNN, multistage PCE attains higher cut values on G1, G2, and G3, whereas QRF-GNN attains higher values on G11, G12, G14, G15, and G16.

Overall, these results indicate that multistage PCE provides a clear improvement over conventional PCE and is competitive with strong classical GNN methods on large Max-Cut instances.
At the same time, the remaining gap on several instances suggests that further improvements in the choice of Pauli correlators, circuit design, optimization schedules, and post-processing may be important directions for closing the gap with the strongest classical methods.

\section{Conclusion}
\label{sec:conclusion}
In this work, we investigated the factors that influence the performance and saturation of Pauli Correlation Encoding through analyses of expressivity and trainability.
First, by comparing PCE with classical surrogates based on tensor networks, we found that PCE achieves competitive solution quality with substantially fewer trainable parameters.
This suggests that its effectiveness is closely related to the efficient propagation of correlations under a limited parameter budget induced by circuit topology.
Second, through supervised optimization toward reference Max-Cut solutions, we showed that even shallow PCE circuits can represent strong solutions.
This result indicates that insufficient expressivity is unlikely to be the dominant limitation in the conventional setting.
Instead, the main bottleneck appears to arise from trainability under the relaxed objective function.

Motivated by this diagnosis, we proposed a multistage continuation framework for PCE.
In this framework, optimization gradually transitions from a smoother relaxed objective function to a sharper objective function aligned with the target discrete problem.
We showed that even the simplest nontrivial case, namely the two-stage setting, already improves performance over conventional PCE.
This improvement is especially clear in the regime of shallow circuits.
Furthermore, the resulting multistage PCE achieved competitive performance with representative GNN methods on G-set benchmark instances with 800 vertices.

Overall, these results suggest that improving the optimization pathway, rather than only increasing circuit expressivity, is a central route to enhancing PCE.
An important direction for future work is the development of a more principled and algorithmic strategy for hyperparameter selection, including the design of the $(\alpha,\beta)$ schedule, the number of stages, and the number of iterations per stage.

\begin{acknowledgments}
  This work is supported by MEXT Quantum Leap Flagship Program (MEXT Q-LEAP) Grant No. JPMXS0120319794, JST COI-NEXT Grant No. JPMJPF2014, JST Moonshot R\&D Grant No. JPMJMS2061, and JST CREST JPMJCR24I3.
\end{acknowledgments}

\appendix
\section{Implementation Details}
\label{app:implementation_details}

% \paragraph{Hyperparameter settings.}
% Unless otherwise stated, we follow the conventional PCE setting of Ref.~\cite{Sciorilli2025Towards} for the hyperparameters $\alpha,\beta$ and normalization$\nu$. In the single-stage setting, we uses $(\alpha,\beta)=(n^{k/2},0.5)$. The normalization factor $\nu$ is chosen so that the regularization term has a scale comparable to that of the main loss term. For weighted Max-Cut, we set
% \begin{align}
% \nu = \frac{W}{2} + \frac{W_{\mathrm{MST}}}{4},
% \end{align}
% where $W=\sum_{(i,j)\in E} w_{ij}$ is the total edge weight and $W_{\mathrm{MST}}$ is the weight of a minimum spanning tree of the graph ~\cite{poljak1986heuristic}. For unweighted Max-Cut, this reduces to the Edwards--Erd\H{o}s bound~\cite{bylka1999maximum}:
% \begin{align}
% \nu = \frac{|E|}{2} + \frac{|V|-1}{4}.
% \end{align}
% Except for the stage-dependent schedule of $(\alpha_s,\beta_s)$, all other components are kept identical to the conventional PCE setting unless otherwise noted.

\paragraph{PI-GNN.}
We implement PI-GNN following Ref.~\cite{schuetz2022combinatorial}.
Specifically, we use a two-layer GCN based on GraphConv units: the first layer maps randomly initialized node embeddings of dimension $d_0$ to an intermediate representation of size $d_1$, followed by an elementwise ReLU, and the second layer outputs a scalar logit per node ($d_2=1$), which is passed through a sigmoid to produce soft assignments $p_i\in[0,1]$.
Following the heuristic in Ref.~\cite{schuetz2022combinatorial}, we set $d_0=\lfloor \sqrt{3|V|}\rfloor$ for graphs with $|V|<10^5$ (and $d_0=\lfloor \sqrt{|V|}\rfloor$ otherwise), and $d_1=\lfloor d_0/2\rfloor$.
For the G-set instances with 800 vertices ($|V|=800$), this yields $d_0=48$ and $d_1=24$.
Training is performed with Adam using a learning rate of $10^{-4}$ for up to $10^5$ epochs with early stopping (absolute tolerance $10^{-4}$, patience $10^3$).
For each instance, we perform 10 independent runs with different random seeds and report the best solution obtained among them.

\paragraph{QRF-GNN.}
We implement QRF-GNN following Ref.~\cite{pugacheva2024enhancing}.
Specifically, we use the default QRF-GNN architecture with recurrent feature updates and two parallel convolutional branches, where the recurrent feature is given by the concatenation of the outputs before and after the sigmoid function from the previous iteration.
For Max-Cut, the dimension of the random part of the input vector is set to $10$, and the hidden-layer size is fixed at $50$.
Training is performed with Adam without a learning rate schedule; the learning rate is set to $0.014$, while the remaining optimizer parameters are kept at their default values.
Gradients are clipped to have a Euclidean norm of at most $2$.
The dropout rate is set to $0.5$.
For the G-set instances considered here, we use up to $10^5$ iterations.
For each instance, we perform 10 independent runs with different random seeds and report the best solution obtained among them.

\paragraph{Multistage PCE.}
We implement the proposed multistage PCE with $K=6$ stages.
The stage-dependent hyperparameters are updated according to a linear schedule from $(\alpha_1,\beta_1)=(n^{k/2},0.5)$ to $(\alpha_K,\beta_K)=(500,0)$.
Thus, for stage $s\in\{1,\dots,K\}$, the parameters $(\alpha_s,\beta_s)$ are obtained by linear interpolation between these endpoints.
Each stage is optimized until the gradient norm falls below $10^{-3}$.
Unless otherwise noted, the same optimization and stopping criteria are used for all benchmark instances considered in this work.
For each instance, we perform 10 independent runs with different random seeds and report the best solution obtained among them.

% \paragraph{Local search.}
% In the local search, we perform a greedy 1-bit-flip refinement by sweeping over all vertices and accepting flips that improve the cut value, iterating sweeps until no improving 1-bit flip exists.
% Using incremental gain updates, flipping vertex $i$ affects only $\Theta(d(i))$ incident edges, so the cost per flip is $\Theta(d(i))$.
% Hence, one full sweep has total complexity $\sum_i \Theta(d(i)) = \Theta(|E|)$, which is linear in the graph size.
\bibliography{cite}

\end{document}